# Risk-Prone and Risk-Averse Behavior in Natural Emergencies: An Appraisal Theory Approach

Sorin A. Matei

Rajesh Kalyanam

Purdue University

Abstract

Individuals who shared actionable information during Hurricane Sandy were significantly more likely to exhibit risk-prone behavior, as measured by a novel *Risk Behavior Quotient* (RBQ). Using a dataset of 36,595 geo-located tweets from 774 users in the New York area, we found that a higher proportion of actional tweets predicted increased exposure to physical even if overall users ultimately moved toward lower-risk zones. This counterintuitive finding suggests that proactivity—manifested in sharing crisis-relevant content—correlates with greater exposure to risk, possibly due to increased mobility or engagement in hazardous areas. In contrast, a greater number of social media peers was associated with reduced risk exposure.

This study builds on *appraisal theory*, which frames risk-related decisions as outcomes of cognitively mediated emotional and rational evaluations. We extend this theory to digital crisis behavior, distinguishing between emotional and actional appraisals expressed via social media. Tweets were categorized using sentiment analysis and semantic classification, enabling the isolation of affective and behavioral signals.

Our methodology combines natural language processing with spatial vector analysis to estimate individual movement paths and risk exposure based on evacuation and flooding maps. The resulting RBQ captures both direction and intensity of risk behavior, allowing us to model how online communication reflects and predicts real-world risk engagement during natural disasters.

*Keywords*: social media, Twitter, risk behavior, natural emergency, crisis communication, appraisal theory

# Risk-prone and Risk-prone Behavior in Times of Crises

Humans interact with the natural environment purposefully, as agents (Bandura, 1989). The purposes themselves can be defined both by reason and emotion (Weber, 1947). While the distinction between emotional and reasoned responses is fundamental, too often social interventions that aim to understand or change how people interact with their environments lose from sight the blending of emotion and reason in the purposes that move human behavior. Research and practical applications that aim to change human behavior may use a variety of assumptions about human behaviors, but many are often shaped only by reasoned action models (Ajzen, 1985; Lindell & Perry, 2012). Although in theory such models do not exclude emotions and decisions that fork along rational and non-rational choices, they do privilege a linear, reasoned approach.

Building on appraisal theory (Lazarus, 2006; Lazarus & Smith, 1988), we propose a dual emotion-reason model for human decision making in emergency situations. The model is tested in online communication situations to better understand risk behavior. The model focuses on purposeful reactions and their connection to risk. Purposes can be of many kinds, rational or not, value laden or instrumental. The means used to achieve the purposes can also be shaped by a variety of instrumental or emotional choices. In situations of danger, crisis, or uncertainty, human responses vary as a function of human purposes (Lazarus, 2006), some of which are directly, while others only indirectly related to immediate individual survival or “needs.” Saving or protecting other, more vulnerable members of the family or community, a psychological need to find refuge in a familiar place (Fullilove, 1996), rather than to flee into the unknown, incomplete knowledge or wishful thinking, or expectations to beat the odds (Johnston, Bebbington Chin-Diew Lai, Houghton, & Paton, 1999; Pender, 2001), all can come into play when a decision needs to be made in times of crisis. Furthermore, in situations of acute stress individuals might feel the need to express and share emotions as a means to compensate for the lack of control over their circumstances (Lazarus, 2006).

Risk-related decision-making in emergency situations is influenced by many inputs, some accurate, some not. Many come with subjective assumptions, some valid, and some not. Across situations, reason and emotion will be expressed in variable amounts, too. If we want to explain human behavior in crisis situations we need to ground the research on the sound basis of a theoretical framework for human choices formulated in the context of acts driven by purposes. We use as a lens the psychological model of human response to stress situations derived from appraisal theory (Lazarus & Smith, 1988), which emphasizes the need to examine both emotional and reason-inferential responses. We also need to distinguish between rationality with respect to purposes, and rationality with respect to means (Weber, 1947). Purposes, again, cannot be seen as automatic triggers. They cover a broad array of existential needs, which are modeled by many possible constraints, such as norms, ethical injunctions, emotional reactions, or rational interests. In times of predictable natural emergencies, such as floods, storms, or tornadoes, human responses to danger and reactions to calls to action are rarely purely instrumental or driven by reason (Shklovski, Palen, & J Sutton, 2008). If it were so, we would expect sudden and unidirectional rushes away from danger. Similar processing mechanisms, uniformly rational or instinctive, lead to similar responses. Yet, this is not what we see in human responses, which are more diverse. This diversity is, at the same time, explicable if looked at from the perspective of cognitive psychology.

Appraisal theory: tenets and applicability to crisis responses

Appraisal theory is a general cognitive theory of emotions (Lazarus & Smith, 1988). Emotional responses are seen as complex cognitive processes, which build on, but go beyond biological reactions to environmental signals. The core proposition of appraisal theory is that emotions are responses elaborated cognitively, by which individuals satisfy the fundamental need of environmental surveillance. Emotions are appraisal responses. Individuals appraise their environs continuously and the emotional elaborations of their appraisal serve the need to understand and cope with the environment. Appraisal emotions may in turn lead to further actions and interactions, which may alleviate the emotional state or enhance it. Appraisal itself estimates the impact of environmental stressors on the individual, the likelihood that the impact will affect individual purposes and acts, and the ability of the individual to control or not the entire situation or to react to it in a given way.

Appraisal theory is not only concerned with emotion. It also allows for problem-focused, actional responses. Emotional responses are a type of action themselves, but the response is cognitive and at best expressive. When individuals are scared, they experience fear and express the fear through messages and certain expressive acts. However, the same signal that triggered the fear response can lead to planning and executing reason-driven actions. Sure, the reactions might be rash or ill-considered, but it is not the validity or justification of the action, but the reasonable expectation of the actor that certain tangible activities might lead to specific outcomes that might remove the environmental stressor.

Critical situations, such as exposure to danger or threat of danger during weather or natural emergencies, inevitably trigger appraisal mechanisms, which may lead to either emotional or actional responses. While both responses are coping mechanisms, in certain situations they may work together while in others at cross-purposes. Furthermore, emotional responses may filter and direct the actional responses in certain ways. The interplay of the two responses will affect in a direct way the likelihood of individuals to perceive and to expose themselves to risk. The manner in which emotional vs. planned action reactions to natural emergencies intersects with risk behavior is, however, insufficiently known. In fact, most literature on human behavior during natural emergencies deals with what people say or think, rather than what they actually do.

As a theory of purpose-driven evaluation of environmental stressors, appraisal theory lends itself quite well to understanding risk behavior. Individual risk-prone or averse behaviors are the product of evaluations, some of them emotional and some of them actional. The core question in this context is: what leads to risk-prone or risk-prone behaviors? Although at first sight, one might be tempted to believe that actional responses might lead to risk aversion, while emotional ones to risk-prone behavior, the situation could be quite the reverse. In the first scenario, where actional responses prevent risk, the presumption is that actional responses are more deliberate reason based. This, however, ignores the possibility that action responses are by definition engaging. Actional responses compel the actor to act, to move, to do something. On the other hand, emotional responses may or may not lead to tangible actions. Emotional responses, such as fear or anxiety, can in certain conditions dampen tangible reactions, expending themselves through introspective reflection.

Since to our knowledge appraisal theory has not been for understanding risk-prone or risk-prone behavior, we propose the following exploratory research question:

RQ1: Which risk behaviors (actional vs. behavioral) are associated with emotional or actional reactions to natural emergency situations?

Responses to danger and their actional or emotional elaboration do not, however, take place in a vacuum. Each individual action, especially risk behaviors, takes place on the backdrop of other actors' actions. Decisions to act or to emote are, as well, influenced by group reactions and interactions. In determining the effect of responses to natural emergencies on risk behavior we need to consider the possibility that the responses of the peers might amplify, modify, or minimize the reactions to danger and subsequently risk behaviors. Again, since we do not have an extended appraisal theory for group effect on individual appraisal, we can only propose research questions to consider the effect on self vs. peer emotional or actional responses on risk behavior. Thus, a second research question is:

RQ2: Do peer emotional or actional responses influence risk behaviors in natural emergency situations?

## Methodological Approaches to Quantifying Risk and Responses to Risk in Natural Emergencies

Building on previous work on spatial behavior in times of natural emergencies (hurricane flooding), we propose a research strategy that captures emotional, actional, and risk behaviors from social media streams. First, we mine social media content to capture emotional or actional appraisal of natural emergencies. Second, we observe the travel behavior reflected in the same social media streams to quantifying exposure to risk and by this inferring risk-prone or risk-prone behaviors.. The analytic strategy relies on harvesting from social media, at this point

Twitter, those objective records that capture human cognitions and spatial behavior. The data is extracted both from the the textual content of the social media flows as well as from metadata associated with the posts, especially the time and the location of the posts. The social media content is processed in two ways. First, it is geocoded to obtain data points for mapping exposure to risk and quantifying individual risk-prone behavior. Second, it is categorized using a taxonomy of goals and responses divided, according to appraisal theory, along the rational-emotional divide. The resulting categorization helps us model at the individual and group level the responses that moderate human behavior in times of crisis. In what follows, we will present the geocoding and risk behavior quantification process, followed by the methodology for categorizing social media content for emotional vs. actional responses. The procedures are deployed within a pilot study that captures travel and risk behaviors in a crisis situation (hurricane Sandy). Behaviors are explained on the basis of actional and emotional responses detected in the same tweets that captured risk behavior.

## Quantifying Behavioral responses to Crisis and Measuring Exposure to Risk

Critical situations, such as natural emergencies, involve exposure to physical harm. Travelling in a natural emergency exposes the individual to various degrees of risk. Cumulative analysis of risk exposure can tell us if an individual was more or less risk-prone during a given emergency situation. To capture risk behavior, we need to track the movements of the individuals and then quantify the degree to which they have been exposed to risk by assessing the time and distance covered while in dangerous situations throughout all the areas he or she has travelled through.

To characterize the actual risk-prone or risk-avoidance behavior of each individual we use their social media (Twitter) output. We start with geo-tagged tweets or posts. Each is placed on a timeline and the path of the user is described as a chain of segments. For each segment we know the distance travelled, the speed, and more important the heading. This is used to calculate the average heading of the entire path, using methods derived from vector algebra. The result is a vector characterized by direction (average heading) and intensity. Average heading is calculated using a methodology derived from vector algebra, as utilized in the environmental sciences. Specifically, we used the Meteorological Resource Center formula for calculating average vectors, as developed in the context of averaging wind directions (http://www.webmet.com/met_monitoring/62.html). The source methodology is appropriate in this context, as the basic problem is almost identical, namely, determining an average heading and intensity for a set of vectors.

We start with N observations of $\theta_i$ (azimuth angle of the vector, measured clockwise from the north) and $u_i$ (magnitude, measured in terms of speed). According to the Webmet methodology, the mean east-west, $V_e$, and north-south, $V_n$, components of the geographic vectors are:

$$V_e = -\frac{1}{N}\sum u_i \sin\theta_i$$

$$V_n = -\frac{1}{N}\sum u_i \cos\theta_i$$

$$\overline{U_{RV}} = \sqrt[2]{\left({V_e}^2 + {V_n}^2\right)}$$

$$\overline{\theta_{RV}} = \tan^{-1} {V_e}/{V_n} + FLOW$$

Where,

$$FLOW = +180; for \ \tan^{-1} {V_e}/{V_n} < 180$$

$$FLOW = -180; for \ \tan^{-1} {V_e}/{V_n} > 180$$

Using the resulting vector, we calculate the corresponding risk behavior of the user. This is the difference between the level of risk at the end point of the average vector, minus the value of risk at the origin, plus the value of risk at the end point, multiplied by distance and speed travelled.

$$RBQ = \left((R_d - R_o) + R_d\right) \times (d \times s)$$

Where,

*RBQ=Risk Behavior Quotient*

$R_d$ - Risk level at destination

$R_o$ - Risk level at origin

$d$ - distance

$s$ – speed

Risk at destination point is present twice to account for the fact that if the user is stationary (starts and ends up in areas with the same risk values or does not move at all) the “0” difference does not suppress the fact that the user was, after all, in an area of certain risk level. Multiplying by speed and distance, aims to take into account both velocity and effort implied by distance in determining the “load” of the risk behavior. Two individuals who travel from the same level of high to low risk areas will display different levels of risk behavior quotient if they travelled further or expended more time travelling. Lower RBQ scores indicate less risky behavior.

Figure 1 illustrates the methodology, using as examples geo-tagged tweets posted by two theoretical Twitter users during Hurricane Sandy (2012). The New York City metropolitan area is divided into the actual evacuation/risk areas used at the time by the city emergency management office. The redder the area, the more urgent to evacuate and the riskier for the population to stay in. We allocate to each area a score: A (red), riskiest, 4; B (orange), moderate risk, 3; C (yellow), low risk, 2. Non-colored areas have a risk score of 1. User 1 travelled 6 miles in one hour, tweeting twice. The heading of his vector of risk behavior corresponds with that of his movement, taking him from a 2 to a 1 zone. Thus, he displays a behavior that mitigates risk. He tweets once in the moderate risk area of Queens, score 3, and once in Jamaica, a low risk area with a score of 2. His risk behavior will be the product the difference between the value of risk at the destination and that at origin added to that of the area of destination, multiplied by distance, and speed. For user one RBQ = (2-3+2) *6*6=36. The second person tweets the first time in Queens, in an area of moderate risk. She ends up in an area of safety, also Jamaica, close to the first user. She travels a total of 22 miles, over a period of an hour, at an average speed of 20 miles an hour, tweeting 7 times. RBQ quotient for this person gives (2-3+2) *22*20=-440. As we can notice, the RBQ for user 1, who travelled less and moved from danger to relative safety, is much smaller than that of the second person, who although has started and ended up in the same areas of risk, has spent more time traveling, exposing her to maximum danger at one point.

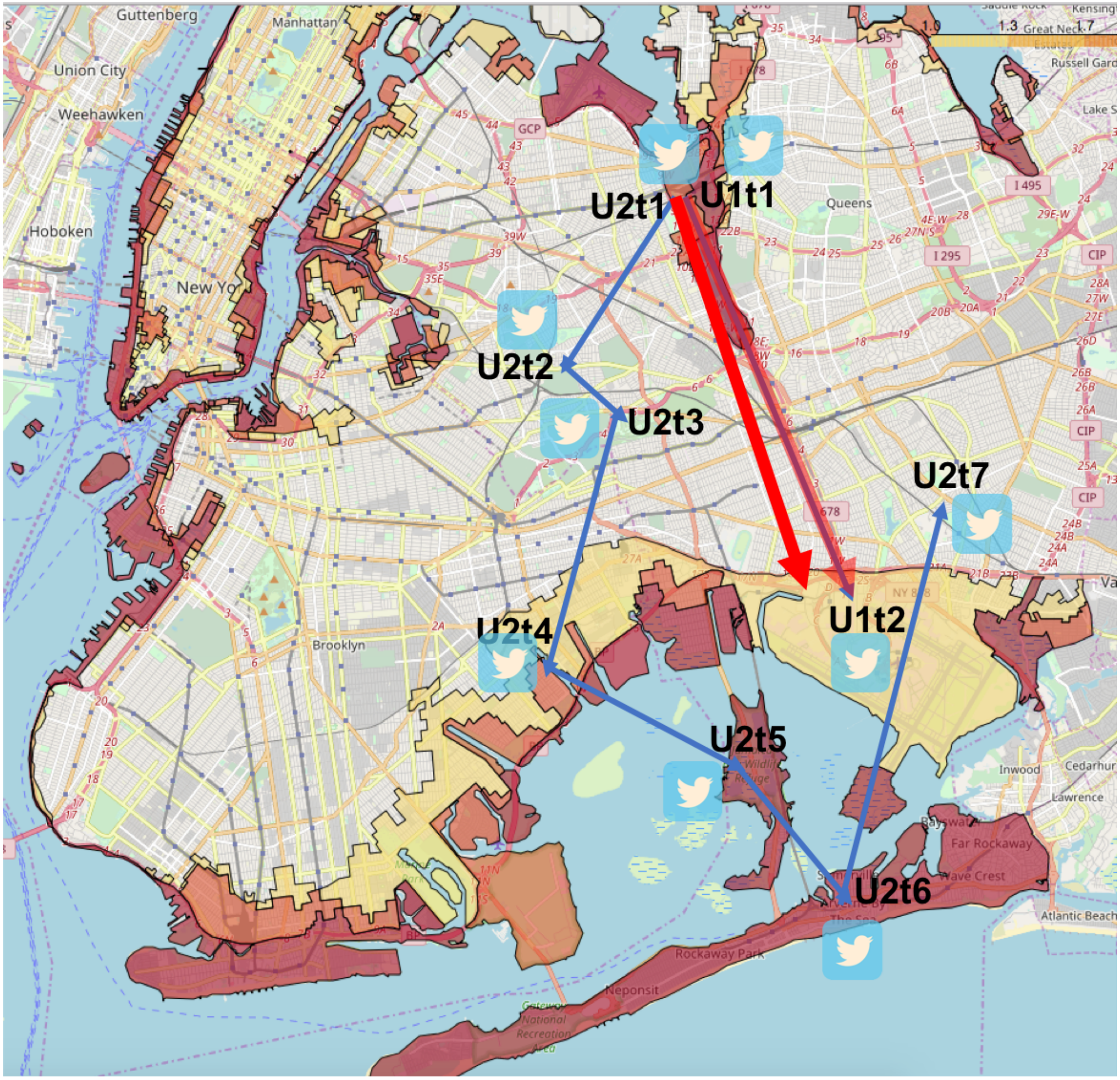

*Figure 1: Map illustrating the movement and tweeting behavior of two theoretical users (U1 and U2). Blue arrows show the actual path of movement, red indicate the average vector.*

The RBQ score is of crucial importance in the context of this type of research. It captures not only where some individual travels, but more importantly his or her implicit attitude toward risk. RBQ captures to what degree the person tended to run away, go toward, or stay relatively stationary in a time of crisis. This can then be explained as purposive action, when regressed, with adequate controls, on the taxonomy of appraisals captured from the social media feed using statistical content analysis and natural language processing. Specifically, the methodology indicates if the movement and the underlying attitude is explained by specific purposes, by constraints, or by other factors, controlling for each other. The expectation is that risk behavior quotients will load negatively or positively with specific purposes.

**Categorizing Social Media Content for Appraisal Responses during Natural Emergencies**

Once we capture through social media content mining the degree of risk entailed in the travel behavior of an individual caught up in a natural emergency situation, we analyze the content itself to capture the appraisals predicted by appraisal theory. Each individual tweet used for calculating RBQ is categorized as an emotional vs. actional response. For the purposes of this pilot study, response differentiators will be semantic and sentiment analytic. Specifically, we classify each unit of social media content as “actional” or not utilizing three separate semantic textual analyses. Word tokenization is first used to identify the verbs in each tweet, only considering a certain set of "strong" verbs (of all tenses) including "do", "go", "say", "watch", "want", and "need". Tweets utilizing these verbs are considered "actional". Since we also consider the act of providing information about the event as "actional", we identify neutral tweets that mention hashtags relevant to the emergency. In this case of Hurricane Sandy, these include "Sandy", "storm" or "hurricane". Furthermore, we use bag-of-words topic modeling to extract the set of top four topics characterizing tweets from official accounts such as the New York Governor's office, FEMA, nycem and the National Hurricane Center during this period. Our reasoning behind this is that these tweets are highly likely to be informational in nature. We then use topic classification to identify the likelihood that a user's tweet matches these topics. If the total likelihood across the four topics exceeds 0.66, we consider that tweet to be "actional". We can then derive an "actional" score for a user based on the proportion of their tweets considered "actional" to their total number of tweets.

The identification of neutral tweets relies on sentiment analysis, which we also use in characterizing emotional responses. The Stanford CoreNLP software is used to extract a set of five scores for a tweet's text ranging from very negative, negative, neutral, positive and very positive. These five scores are then converted into three by merging the very negative and negative scores and the very positive and positive scores respectively. A threshold score of 0.66 is then used to classify a tweet into exactly one of these three classes.

**Pilot study**

We used 195,474 Twitter status updates issued in the New York and New Jersey area to assess the feasibility of using social media content for measuring exposure to risk during a natural emergency: hurricane Sandy (October 28 - October 30, 2012). Of these, however, we further refined the dataset by only including 36,595 tweets. The selection process involved a multi-step procedure of identifying the most relevant topics and tweeting users for the given crisis situation. First, we discard any users who did not tweet from at-least two separate locations. Since our analysis is centered around risky travel behavior during a natural emergency, we are only interested in users who moved around during this event. Next, we only retain users who have mentioned the keywords "Sandy", "at", "go", "drive" or "driving" in at-least one tweet. This decision was justified by the requirement to restrict our analysis to those users who were tweeting about the event in question and including some "actional" information in their tweets. Finally, we restrict our analysis to only users originating in the five New York boroughs. Since we currently only have information on the evacuation zones and flooding of these boroughs, our risk analysis is restricted to these users alone. At the end, we examine 774 users and the 36,595 tweets by these users.

The tweets were issued right before and during the landfall of the hurricane on October 28, 2012. The first recorded tweet was issued at 2:30 pm on 10/28/12 and and the last at 4pm 10/31/12. As is known, the hurricane was supposed to and did indeed affect the New York City area through flooding, rain, strong winds, and lightning. The event was intensely publicized and has affected the residents significantly.

Each tweet was geolocated at a certain point on the map and the individual who issued it was assigned, for that specific point in time a given level of risk exposure. Risk exposure for each area of the map was calculated using two "ground-truth maps:" the pre-emptive evacuation map issued before the incident by the New York City government (https://project.wnyc.org/hurricane-zones/) and one of actual impact, compiled after the event by the National Weather Service (https://project.wnyc.org/flooding-sandy-new/index.html#9.00/41.3601/-71.3739 and https://www.weather.gov/okx/HurricaneSandy ). Overlapping the two maps, we obtained a "risk map" (Figure 3), whose values span from intense risk (area was both pre-emptively and post-factum found to be flooded and affected by the storm at the highest level) to no-risk (area was neither expected or actually flooded or affected by the storm). The gradations of risk of different regions was assigned based on these two maps. Each evacuation zone corresponds to a risk level, with the area of highest risk at risk level 3, followed by 2, 1 and 0 assigned to areas that do not fall in any evacuation zone. If a location was determined to be flooded this risk level was increased by 1. In effect, this gives us a gradation of risk levels ranging from 0 through 4. We illustrate the overlays of the evacuation zones and flooded regions in Figure 3.

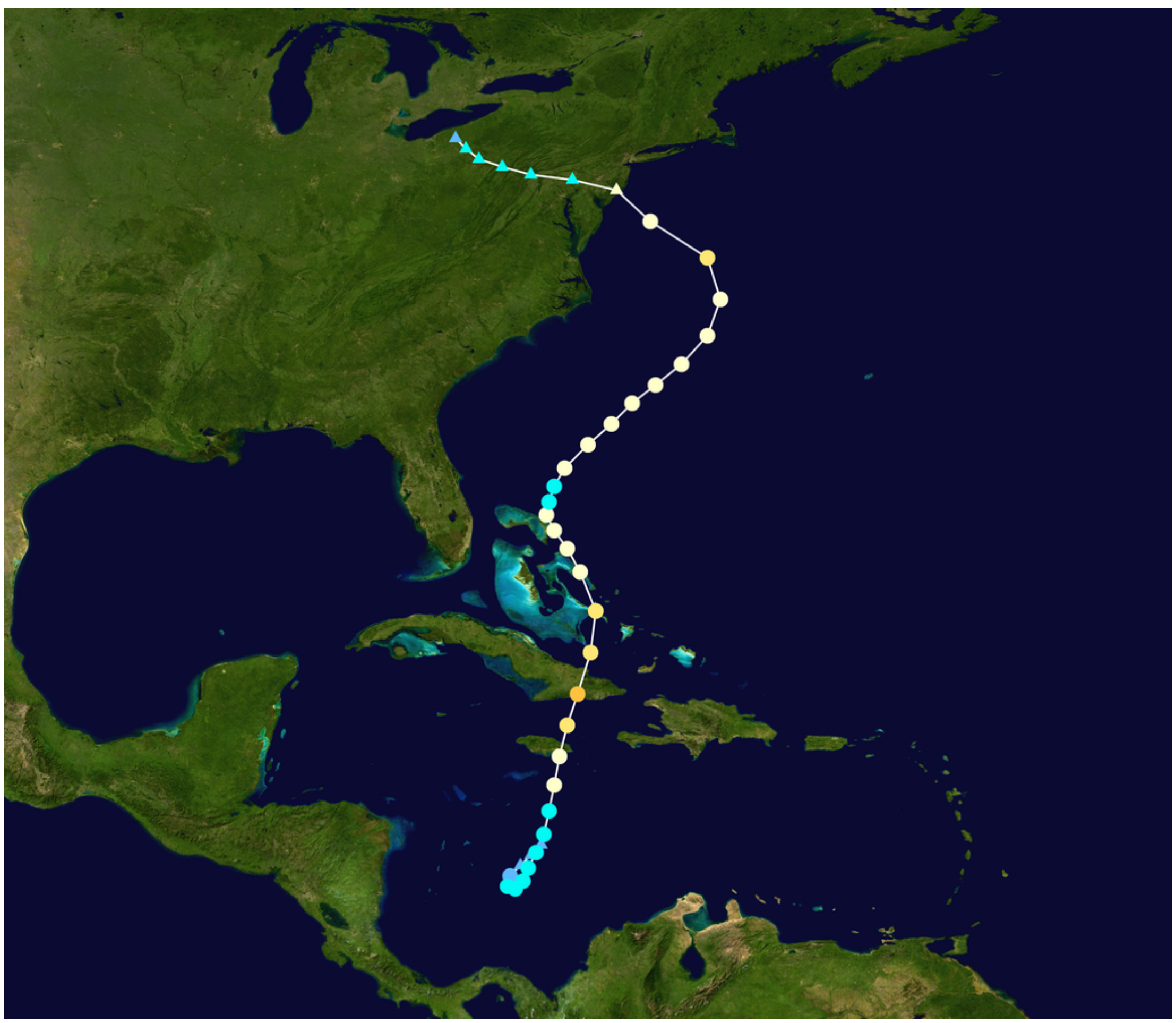

*Figure 2 : The path of Hurricane Sandy*

Once each tweet was assigned a level of risk exposure, a total risk quotient was calculated for each individual. The procedure, described in detail above, used vector analysis to calculate the average movement vector bearing and length. The final quotient for each individual

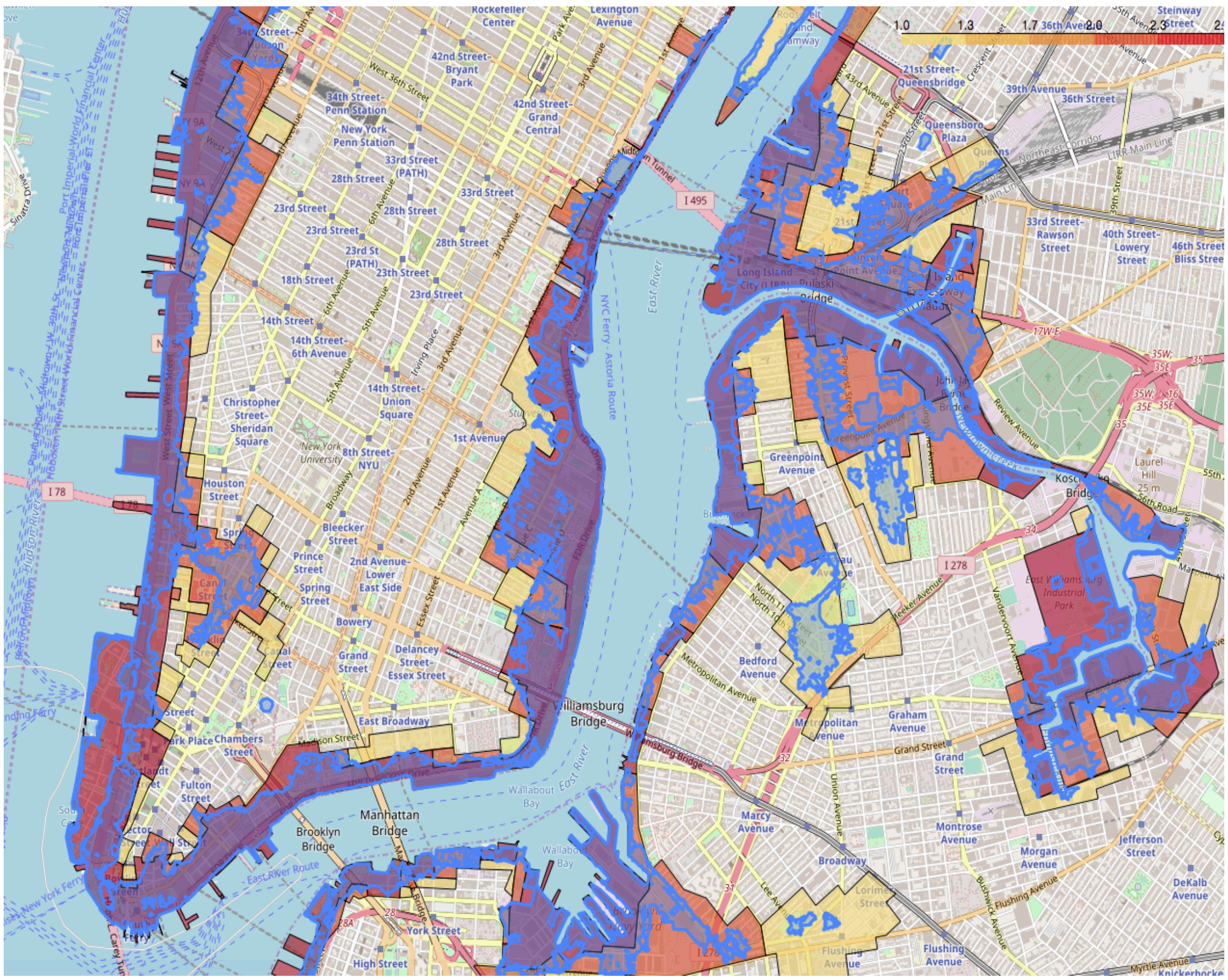

*Figure 3: A map of evacuation zones for New York (color coded in red, orange and yellow indicating highest to lowest risk respectively) overlaid on a map of the actual flooding during Hurricane Sandy (in blue)*

was obtained by subtracting the risk exposure value of the journey origin from destination point, as identified by the average vector. Significantly, the vector destination point takes into account the circuitous trips the user took, which lengthened and shortened by the time elapsed. If the difference between the destination and origin point is positive the user was more likely to be exposed to risk and if the difference was negative, he or she was less likely to be exposed to risk.

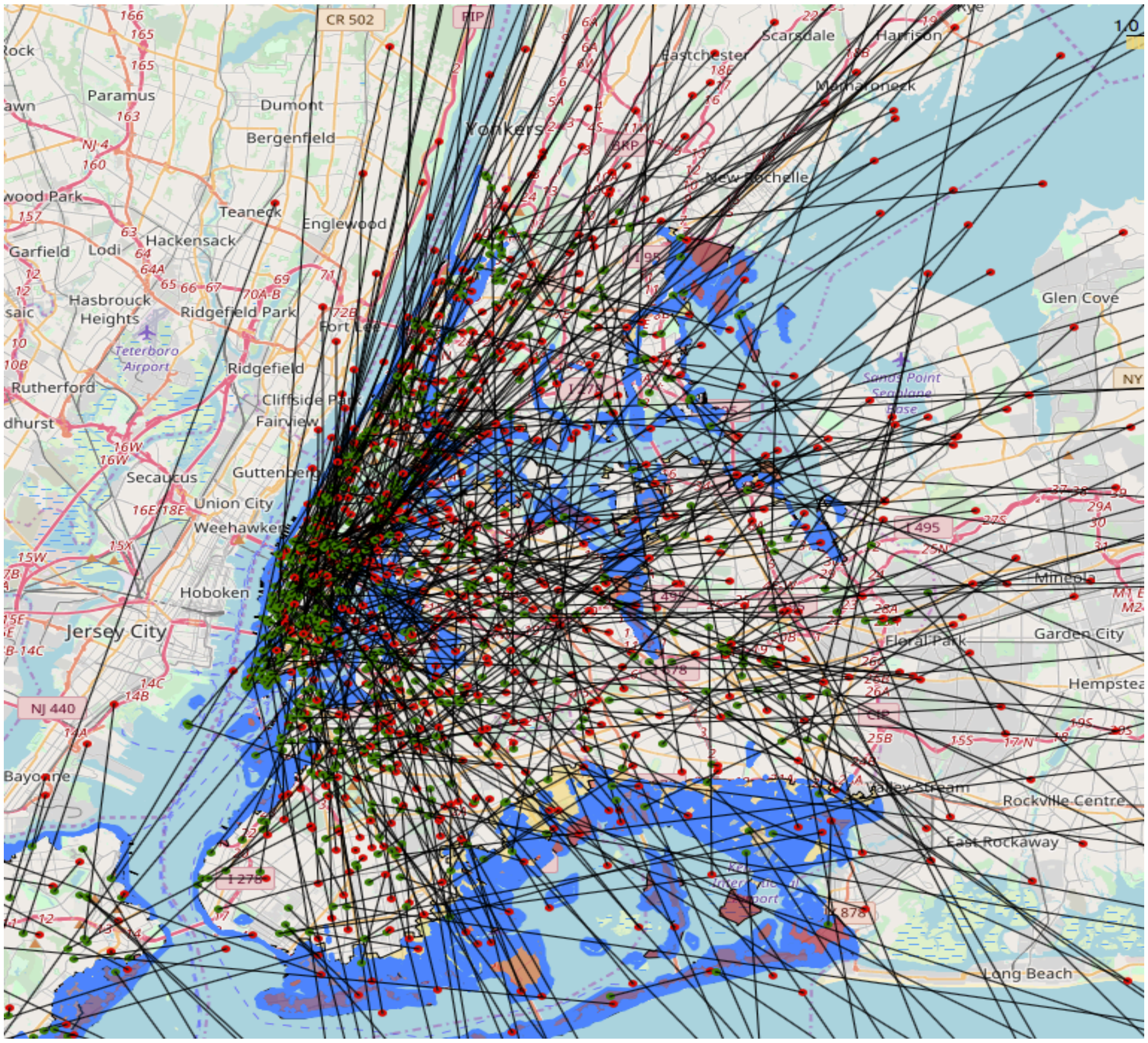

*Figure 4: A map of New York depicting the average travel vectors of all the 774 users from our study. The origin points are in green and destinations in red*

The map in Figure 4 visualizes the individual average vectors. As can be seen, at first sight the movement might look quite Brownian. However, our risk assessment methodology can be enhanced with a group level calculation of an average vector of risk for all individuals, taken together. Using the same individual methodology for averaging individual step-by-step vectors, we can average the final individual average vectors for each individual to obtain one final average origin and destination point for all the individuals observed for this study.

The map to the left in Figure 5 shows the locations of these two average points. The visualization shows in telling manner, a simple story. On average, the individuals that tweeted during hurricane Sandy moved to safer areas, since the red (destination) marker is in a safer area than the average green (origin) marker. Specifically, the average of all origin points was [40.74, -73.95], which is low to moderate risk, while the location of the overall destination point was LOWER, at [40.75, -73.86], indicating lower risk.

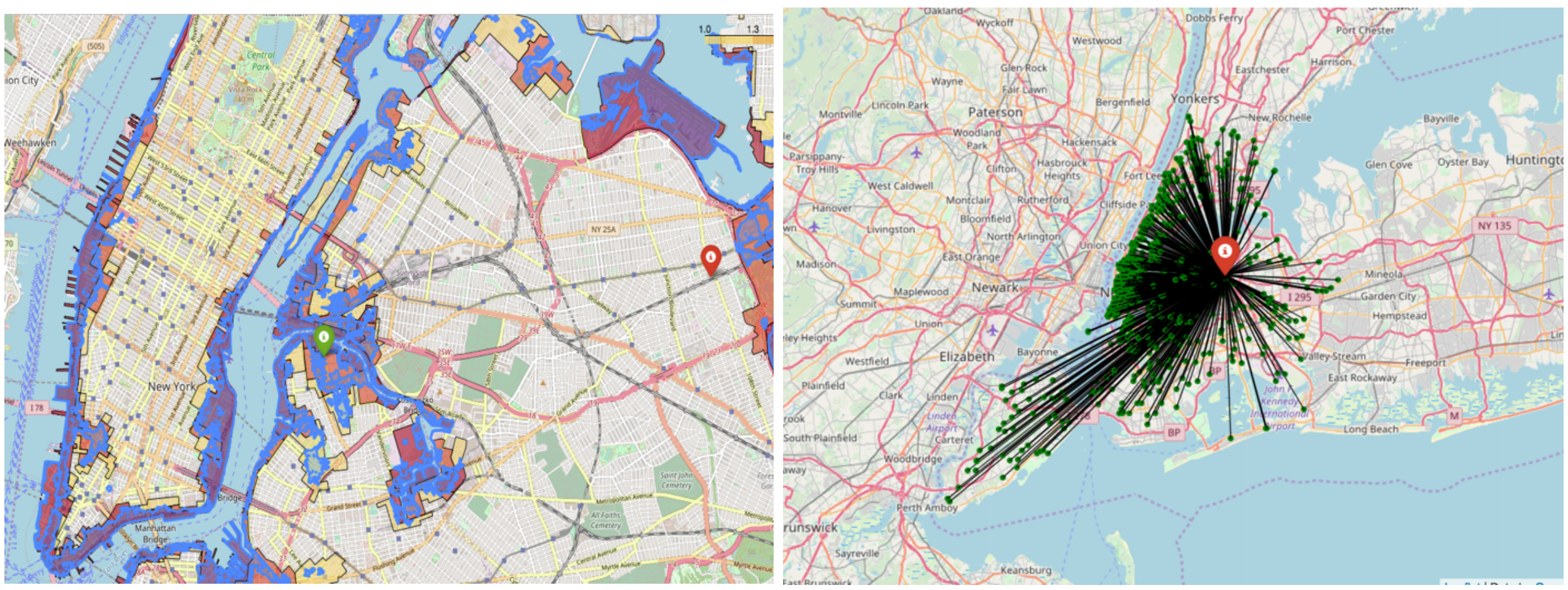

*Figure 5: A map of New York depicting the overall average origin (in green) and destination (in red) points (left) and the position of the average destination point relative to the departure points of all the Twitter authors (right).*

However, this is only part of the story. Our risk quotient also captures overall individual risk exposure scores due to exposure during travel. These vary between -703 and 5687. More important, the average value is 12.2 and the median 0.18. Being positive, this indicates that although on average individuals ended up in a safer spot than where they started, their exposure to risk was, although low, positive because throughout their travels they were significantly exposed to danger. Most individuals included in the study were risk-prone, not risk-averse, although only marginally so.

The second step in our analytic process was to address the research questions by relating the individual risk quotient to the users' own and to their peers' actional and emotional responses. The independent variables were calculated by categorizing the tweets, counting them and then by calculating the proportion of tweets in the emotional vs. actional categories out of the total number of tweets. Peer-level variables were calculated by averaging the same values (counts followed by ratios for actional and emotional responses) for the group of users any given user interacted with by tweet answer (reply) or tweet mention (@). Table 1 presents the variables entered in the model.

Table 1. Variables entered in the model

| | N | M | SD |
|---|---|---|---|
| Risk Quotient | 774 | 12.21 | 230.24 |
| Total number of self tweets | 774 | 47.28 | 69.95 |
| Proportion self informative (hashtag) tweets | 774 | 0.16 | 0.22 |
| Proportion self actional tweets | 774 | 0.21 | 0.12 |
| Proportion self emotional tweets | 774 | 0.0010 | 0.0064 |
| Number of peers | 774 | 14.88 | 21.42 |
| Total number of tweets by peers | 774 | 558.01 | 806.66 |
| Proportion peer informative (hashtag) tweets | 774 | 0.08 | 0.10 |
| Proportion peer actional tweets | 774 | 0.07 | 0.09 |
| Proportion peer emotional tweets | 774 | 0.0004 | 0.0012 |

Table 2. Variables predicting the Risk Behavior Quotient

| Variable | Beta | Sig. |
|---|---|---|
| Number of peers | 0.97 | 0.001 |
| Total tweets issued by peers | -0.49 | 0.005 |
| Proportion of actional tweets by user | 0.1 | 0.002 |
| Proportion of tweets by user referring the storm | 0.138 | 0.001 |

We used the SPSS automatic linear modelling procedure to find the best fit model. The final model, with a small r-square=.04, presented in Table 2, retained as significant predictor variables total number of peers, total amount of tweets issued by peers, proportion of actional tweets and frequency of mentioning the crisis situation by hashtags. More significantly, risk exposure increased with high rates of tweeting actionable information, higher rates of mentioning the crisis situation (storm specific hashtags) and total number of tweets issued by the peers. On the other hand, exposure to risk decreased, all things being equal, with total number of peers.

Thus, the answer to the two research questions can be formulated as follows.

RQ1: Which risk behaviors (actional vs. behavioral) are associated with emotional or actional reactions to natural emergency situations?

Actional responses are associated with risk-prone behaviors. Also, discussing and referring to the context of the action is associated with risk-prone behviors.

RQ2: Do peer emotional or actional responses influence risk behaviors in natural emergency situations?

The types of tweets issued by the peers do not influence risk. However, the number of peers as well as the amount of content they put out does matter. Individuals with a wider social network are more risk-prone, while those that are surrounded by a more active network of peers, which produces more content, are more risk-prone. Let us now discuss the implications of these findings.

## Discussion

The present paper uses appraisal theory to assess the impact of emotional vs. actional content on risk behavior on natural emergencies. It also proposes a new methodology to assess human exposure to risk in natural emergencies by using social media (Twitter) content.

We used a two-step analytic strategy. First, we constructed an innovative method to measure risk and validated it. We showed how a new type of vector analysis may reveal fine distinctions in risk behavior during natural emergencies. We detected a small but positive propensity for exposure to risk. However, the vector analysis indicated that although the individuals studied ended up exposing themselves to risk to a certain degree, the end point of all the movements was in an area of slightly lower risk than that of the average origin points. The distinction might sound technical, yet it points to the explanatory power of our approach, which can be used to assess both the global process of exposure to risk as well as the specific topology of the risk landscape in times of natural emergency. When averaging the values for exposure to risk we obtain the overall, over-time exposure to danger. This procedure is similar to other measurements to exposure to environmental harm, such as amount of exposure to UV radiation or second-hand smoke. When we calculate the average vector destination point, by taking into account each individual movement vector, we focus on the outcome, on the product, not on the process of exposure to risk. In this case, the outcome was a general movement from higher to lower risk. In other words, we can say that although during the study period the users tended to expose themselves to risk, they did it for a purpose, which was that in the end the move to a more sheltered location.

The second step of our study aimed to discern the possible relationship between emotional or actional responses to the crisis situation on exposure to risk. We proposed two research questions, one asking if emotional vs. actional content may impact risk behavior and if the content of one's peers may also do the same. To this end, we categorized the same tweets used to quantify risk into emotional (using sentiment analysis) and actional (using semantic analysis) responses. Regression analysis showed that an increase in actional tweets tended to increase exposure to risk, as did a higher volume to tweets issued by one's peers and frequent references to the crisis situation, as indicated by usage of crisis specific hashtags. In other words, we found that a pro-active attitude, referring to actions and to the situation at hand is accompanied by a general tendency to travel around and expose oneself to risk. A first explanation could be that hyper-active users were at the same time high-risk individuals. A second conclusion could also be that in times of crisis we should not try to over activate the users, by asking them to act or to ask others to act just for the sake of acting, as these activities can lead to higher exposure to risk.

On the other hand, a higher number of peers seems to mitigate exposure to risk. This can be explained by a social support perspective. Those that interacted more with other individuals, for whatever reason, were more likely to expose themselves to risk less.

Although these findings might sound intriguing, we need to use caution in interpreting these pilot testing results. The data was limited to a small section of the individuals caught up in the crisis situation. Geocoding was limited to available data, which in the context of Twitter research could be quite noisy. Due to the fact that we could only locate tweets that were explicitly geocoded by the users, we had to look at only a very small proportion of tweets involved in the crisis situation. The methodology for identifying emotional vs. actional content tended to over-simplify the recognition patterns. Finally, the total amount of variance explained by the regression models is rather low, indicating either a noisy or underspecified model.

Future research should use a far more sophisticated data collection method, which should include a more comprehensive sample of individuals, more clearly and reliably tracked as they move in space and time. An alternative method would be to examine another medium, such as Facebook, which more recently has introduced the “check in” methods for natural disasters and which also allows more clear geocoding of status updates.

Future research should also use more sophisticated natural languages processing for determining the content of the tweets. A more mature methodology, which would look at semantic, rhetorical, and sentiment analysis could more directly and reliably identify the nature of the intentions expressed in social media content.

However, despite limitations, the current study is a first pilot attempt to propose a tangible methodology for studying extremely important issues, namely risk and motivation for exposure to risk in natural emergencies. We hope that in the future we will be able to further develop it and create a more comprehensive and reliable research framework for this emerging area of research.